\documentclass[aps,final,notitlepage,oneside,twocolumn,nobibnotes,nofootinbib,
superscriptaddress,noshowpacs,centertags]{revtex4-1}

\usepackage[utf8]{inputenc}
\usepackage[english]{babel}
\usepackage{graphicx}
\usepackage{latexsym}
\usepackage{amssymb}
\usepackage{amsmath}
\usepackage{float}

\begin{document}

\title{Gravitational waves from the merger of two primordial black hole clusters}

\author{Yury Eroshenko}\thanks{e-mail: eroshenko@inr.ac.ru}
\affiliation{Institute for Nuclear Research of the Russian Academy of Sciences, Moscow 117312, Russia}
\author{Viktor Stasenko}\thanks{e-mail: vdstasenko@mephi.ru}
\affiliation{Institute for Nuclear Research of the Russian Academy of Sciences, Moscow 117312, Russia}
\affiliation{National Research Nuclear University MEPhI, Moscow 115409, Russia}

\date{\today}

\begin{abstract}
The orbital evolution of a binary system consisting of two primordial black hole clusters  is investigated. Such clusters are predicted in some theoretical models with broken symmetry in the inflation Lagrangian. A cluster consists of the most massive central black hole surrounded by many smaller black holes. Similar to single primordial black holes, clusters can form gravitationally bounded pairs and merge during their orbital evolution. The replacement of single black holes by such clusters significantly changes the entire merger process and the final rate of gravitational wave bursts in some parameter ranges  (with sufficiently large cluster radii). A new important factor is the tidal gravitational interaction of the clusters. It leads to an additional dissipation of the orbital energy, which is transferred into the internal energy of the clusters or carried away by black holes flying out of the clusters. Comparison with the data of gravitational-wave telescopes allows one to constrain the fractions of primordial black holes in clusters, depending on their mass and compactness. Even the primordial black hole fraction in the composition of dark matter $\simeq1$ turns out to be compatible with LIGO/Virgo observational data, if the black holes are in clusters.
\end{abstract}

\maketitle 


\section{Introduction}

The possibility of primordial black hole (PBH) formation in the early Universe was predicted theoretically in the work \cite{ZelNov66} and then from other points of view it was discussed in \cite{Haw71}. In the work \cite{Car75}, a model of the PBH origin from adiabatic density perturbations was developed in details. Later, other models of the PBH formation during early cosmological dust-like stages \cite{KhlPol0}, as a result of domain wall collapses \cite{BerKuzTka83,RubSakKhl01,KhlRubSak02}, and from collapsing regions with baryon charge fluctuations \cite{DolSil93} were proposed.

Although the existence of PBHs has not been proven yet, their possible observational manifestations are being studied in various cosmological and astrophysical aspects \cite{Caretal20}. In particular, the possibility remains open that the PBHs make up the entire dark matter (hidden mass) in the Universe, if their masses lies in the mass window  $10^{20}-10^{24}$~g \cite{CarKuh20}. Other possible observational manifestations of PBHs include the effect of quantum evaporation of low-mass PBHs with radiation emission, radiation from accretion on the PBHs, as well as various types of dynamic influence of PBHs on stars, globular clusters and galaxies. 

In recent years, PBHs have attracted renewed attention due to the fact that the mergers of PBHs in pairs can explain some gravitational wave bursts recorded by LIGO/Virgo interferometers. These are mainly those bursts in which the angular momentum parameter is small. In the works \cite{Naketal97, Ioketal98, Sasetal16}, it was shown that PBH randomly distributed over space can form binary systems if they are close enough to each other. The minor semi-axis of the orbit of such a system is determined mainly by the tidal gravitational interaction with the third closest PBH. The orbit of the PBH pair is compressed under the influence of gravitational wave radiation, and as a result, the PBHs merge with the generation of gravitational wave bursts. In \cite{Sasetal16}, it was shown that the rate of PBH mergers in pairs can provide statistics of events recorded by LIGO/Virgo, however the fraction of PBH in the dark matter should be $f \lesssim 10^{-3}$. 

In the works \cite{RubSakKhl01,KhlRubSak02}, a model of the PBH formation as a result of collapses of closed domain walls was proposed. Moreover, it turned out that the most likely PBHs did not form as single units, but as members of clusters. Such a cluster consists of the most massive central PBH surrounded by PBHs of smaller masses. Note that clusters formed by such a mechanism are virialized gravitationally bound systems already in the early Universe and they can act as individual objects in various dynamical processes. At the same time, the possibility of PBH clustering is not limited only by virialized compact clusters. In the literature, models have been considered for a slight increase in the PBH number density due to the biasing effects (including effects in generalized theories of gravitation), due to the non-Gaussian density perturbations, or due to the settling of PBHs to the center of the host dark matter halo under the influence of dynamic friction.

If the PBHs are in clusters, then, similarly to single PBHs, such clusters can form pairs and merge when the orbit is compressed. The replacement of single PBHs by the clusters can significantly change the entire process of mergers and the final rate gravitational wave bursts. A new important effect is the tidal gravitational interaction of two clusters. Clusters act tidally on each other and deform. This leads to an additional dissipation of the orbital energy, which is pumped into the internal energy of the clusters, leading to their tidal ``heating'' and expansion. The effect of gravitational shocks on the evolution of a globular star cluster was previously studied in \cite{GneOst99, Gneetal99}. In the case of PBH clusters, many processes have a similar appearance, including adiabatic correction.

In this paper, we do not specify the model of the cluster formation and take as a simple example clusters of PBHs of the same masses with a more massive PBH in the center. The statistics of the  cluster pair formation are the same as in the case of single PBHs \cite{Naketal97, Ioketal98,Sasetal16}, however, the dynamic evolution of the orbit may differ significantly due to the tidal forces. At the initial stage, under the influence of tidal forces, the orbits of binary clusters are compressed and rapidly circularized. At the same time, tidal forces at the circular orbit tend to zero, and further evolution is due to the radiation of gravitational waves. This leads to a decrease in the rate of clusters' merge compared to single PBH pairs of the same masses, however, it leaves a range of parameters acceptable from the point of view of observations. Another possible mechanism for suppressing the merger rate is the destruction or perturbation of PBH pairs in early dark matter halos \cite{Raidal18, Vaskonen19, Jedamzik20}. Taking this into account can provide an even greater decreasing of the merger rate, but in this paper we do not consider such effect. 

The majority of cluster pairs spend most of their evolution time in a wide orbit and do not intersect (the orbit pericenter is much larger than the cluster radius). But at the last stage, when the clusters intersect, they merge quickly. This effect was previously noted in the works \cite{KavGagBer18, PilTkaIva22}, where the PBHs surrounded by dark matter halos were considered. In some respects, these systems resemble the clusters we are considering.  In both cases, tidal gravitational forces act on the systems, if they form a pair. Our calculations confirm that at the last stage after the beginning of the clusters' intersection, their central massive PBHs merge rapidly under the influence of dynamic friction.

\section{Tidal dissipation}

\begin{figure}
	\begin{center}
\includegraphics[angle=0,width=0.5\textwidth]{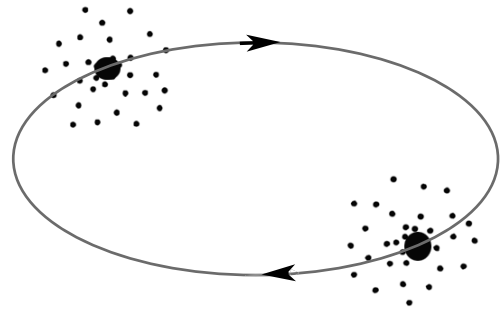}
	\end{center}
\caption{Schematic representation of a system of two PBH clusters with central massive PBHs.}
	\label{grsch}
\end{figure}

Consider a system of two clusters with the same masses $M$ and radii $R$, shown schematically at Fig.~\ref{grsch}. 
Let's assume that the most massive PBH with a mass $M_h$ is in the center of each cluster and is surrounded by $N=M/m$ PBH with smaller masses $m$. We assume that $M\gg M_h\gg m$. This approach is similar to the real model \cite{RubSakKhl01, KhlRubSak02}, when we replace the mass spectrum of the PBH in the cluster with only two masses $M_h$ and $m$. This model is quite realistic, since it takes into account the main dynamic effects, and it is enough to demonstrate the idea of tidal dissipation. In addition, such a model may correspond to the case when a cluster was formed from small PBHs with equal masses, but as a result of dynamic evolution the cluster has a collapsed central core that turned into a single massive black hole. This variant of clusters was considered in the paper \cite{Ero21}.

In the first approximation, the orbit of the clusters is elliptical with large semi-axis $A(t)$ and eccentricity $e(t)$. These values change over time due to the emission of gravitational waves, and under the influence of tidal gravitational forces, which in the general case affect both the orbit of the pair and the internal dynamic evolution (tidal forces ``heat'' the clusters). 

The rate of energy loss in a binary system due to the radiation of gravitational waves is expressed as \cite{LL-2}
\begin{equation}
\dot E_{\rm gw}=-\frac{64G^4M^5}{5c^5A^5}f_1(e),
\label{degw}
\end{equation}
and the change in the orbital angular momentum of the system is 
\begin{equation}
\dot L_{\rm gw}=-\frac{32\sqrt{2}G^{7/2}M^{9/2}}{5c^5A^{7/2}}f_2(e),
\label{lrad}
\end{equation}
where auxiliary functions
\begin{equation}
f_1(e)=\frac{1}{(1-e^2)^{7/2}}\left(1+\frac{73}{24}e^2+\frac{37}{96}e^4\right),
\end{equation}
\begin{equation}
f_2(e)=\frac{1}{(1-e^2)^2}\left(1+\frac{7}{8}e^2\right).
\end{equation}
The eccentricity is expressed in terms of $E$ and $L$ by the following equation \cite{LL-1}
\begin{equation}
e=\sqrt{1+\frac{2EL^2}{G^2M^5}}.
\label{eexpr}
\end{equation}

To calculate the tidal heating of the clusters, we use the formalism from Chapter 3 of the work \cite{Gneetal99}. In our case, the clusters move relative to each other at almost elliptical orbits. Orbits only slowly evolve due to the emission of gravitational waves and tidal dissipation. Let's find an increase in the speed of the PBH with a mass of $m$ inside the cluster. In an elliptical orbit, the velocity is gained mainly during the passage of the periaster, where the cluster experiences a ``gravitational shock''. Using expressions for the Keplerian problem (see \cite{LL-1}) and performing integration along the elliptical orbit by the \cite{Gneetal99} method, we obtain for the average square of the velocity increment for one orbital period 
\begin{equation}
\langle(\Delta v)^2\rangle=\frac{\pi^2r^2GM^2}{A^3m}f_3(e),
\label{dv2}
\end{equation}
where
\begin{equation}
f_3(e)=\frac{1}{(1-e^2)^3}-1,
\end{equation}
and $r$ is the distance of the PBH from the center of the cluster. In this expression, the contribution from the deformation of the cluster in the gravitational field, which is at place even in the case of a circular orbit, is subtracted in parentheses. It can be seen from (\ref{dv2}) that at $e\to 0$, the energy collected during the orbital period tends to zero as $e^2$ (at a fixed $a$). Qualitatively, this behaviour corresponds to the results of numerical simulation performed in \cite{Gneetal99} for the case of globular star clusters, experiencing tidal heating as a result of interaction with large-scale structures of the Galaxy. 

An important role in these processes is played by the so-called adiabatic correction, which depends on the ratio of the characteristic frequency of internal movements of the small PBHs in clusters and the frequency of the orbital motion of the two clusters. The greater this ratio, the less efficient tidal energy transfer is. The adiabatic correction was investigated in the work \cite{Gneetal99} (see also the references in it). The adiabatic correction is defined as the ratio of real energy losses to energy losses in the momentum approximation (when the objects in the cluster are considered to be at rest during the interaction). We will take an analytical approximation for the adiabatic correction in the form of \cite{Gneetal99}
\begin{equation}
K(x)=(1+x^2)^{-\gamma},
\end{equation}
where $\gamma=3/2$, $x=\omega T$, and $\omega\sim(GM)^{1/2}R^{-3/2}$ is the frequency of internal motions.

Taking into account the adiabatic correction, the increase of the energy during one orbital period is
$\Delta E=m\langle(\Delta v)^2\rangle K(x)/2$. Thus, 
\begin{equation}
\Delta E_{\rm tid}\sim\frac{\pi^2GM^2R^2}{2A^3}f_3(e)K(x),
\label{detid}
\end{equation}
therefore the tidal heating of the cluster for one orbital period can be estimated as
\begin{equation}
\dot E_{\rm tid}\sim-\frac{\Delta E_{\rm tid}}{T},
\label{etideq}
\end{equation}
where the orbital period is \cite{LL-1} (the reduced mass is taken into account)
\begin{equation}
T=\frac{2^{1/2}\pi A^{3/2}}{G^{1/2}M^{1/2}}, \quad A=\frac{GM^2}{2|E|}.
\end{equation}
The evolution of the orbital parameters is very slow compared to one orbital period, so the expression (\ref{etideq}) serves as a good approximation for the time-averaged motion.

Let's find the ratio (\ref{detid}) to (\ref{degw}), assuming that the eccentricity is not very small ($e\sim1$)
\begin{equation}
\frac{\dot E_{\rm tid}}{\dot E_{\rm gw}}\sim0.06\cdot(1-e^2)^{1/2}\left(\frac{R}{R_g}\right)^2\left(\frac{A}{R_g}\right)^{1/2}K(x),
\end{equation}
where $R_g=2GM/c^2$ is the gravitational radius of the cluster. From this relation it can be seen that, at least in the intermediate case $1-e\sim1$, the influence of tidal forces on the evolution of the orbit may be more important than the radiation of gravitational waves, and the result may be different compared to the case of a pair of single PBHs. 

Equation for energy loss by the binary system is
\begin{equation}
\dot E=\dot E_{\rm gw}+\dot E_{\rm tid}.
\label{eneq}
\end{equation}
From here and from (\ref{lrad}), (\ref{eexpr}) one can get the following system of equations for $\dot A$ and $\dot e$, 
\begin{equation}
\dot A=-\frac{\pi}{2^{1/2}}\frac{G^{1/2}M^{1/2}R^2}{A^{5/2}}f_3(e)K(x)-\frac{124}{5}\frac{G^3M^3}{c^5A^3}f_1(e),
\label{eqm1}
\end{equation}
\begin{equation}
\dot e=\frac{1-e^2}{2e}\frac{\dot A}{A}+\frac{64}{5}\frac{G^3M^3}{c^5A^4}\frac{(1-e^2)^{1/2}}{e}f_2(e).
\label{eqm2}
\end{equation}
Their numerical solution performed further showed that the main difference compared to the pair of single PBHs lies in the rapid circularization of the binary orbit, and therefore its further evolution occurs due to the radiation of gravitational waves at a circular orbit. Really, due to the symmetry of the tidal forces relative to the movement direction of the clusters along the orbit, the orbital angular momentum is not affected by tidal forces, except for minor second-order corrections. This is also confirmed by the numerical calculation of \cite{KavGagBer18}, where the interaction of PBHs surrounded by dark matter halos was studied. Therefore, the loss of angular momentum is possible only due to the gravitational waves, Eq.~(\ref{lrad}).

In this paper, we consider only clusters for which the time of internal dynamic two-body evolution (approximately 40 relaxation times) exceeds the Hubble time. This situation is the case for sufficiently small masses of PBHs.  In the opposite case, which will be considered in future papers, there are interesting effects associated with the internal evolution of clusters and their mass loss in addition to dynamic evaporation due to two-body relaxation.

\section{The evolution of the orbit and merger rate}

The statistics of the PBH pair formation (without clusters) and the distribution of these pairs by the time of mergers were studied in \cite{Naketal97, Ioketal98, Sasetal16}. At first we are talking about single PBHs in pairs, but then similar equations (except in the case when the cluster size is important) will be valid for clusters. The formation of a gravitationally bound PBH pair occurs at the cosmological stage of radiation dominance at $t<t_{\rm eq}$, where $t_{\rm eq}$ is the moment of matter-radiation equality. The scale factor of the Universe $a(t)$ is normalized as follows $a(t_{\rm eq})=1$, i.e. the scales at the moment of equality are the comoving scales. Then the radiation density is $\rho_r=\rho_{\rm eq}/a^4$.
Let's denote the fraction of the PBHs in the composition of dark matter $f\leq1$, $\Omega_{\rm BH}=f\Omega_m$. At $f<1$, the main part of dark matter is not made up of PBH, but some other objects, for example, new elementary particles. Let us denote by $x$ the comoving distance between some particular pair of PBHs in early epochs, while the orbit of the pair has not yet experienced evolution, whereas the average distance between the components of the pairs
\begin{equation}
\bar{x}=n^{-1/3}=\left(\frac{M_{\rm BH}}{f\rho_{\rm eq}}\right)^
{1/3}.
\end{equation}

According to \cite{Naketal97,Sasetal16}, the minor semimajor axis is determined by tidal forces from the third PBH, located at a comoving distance $y>x$ (i.e. $x$ and $y$ for each specific pair are considered constant). In \cite{Sasetal16} for the large and small semi-axes, it is obtained, respectively, 
\begin{equation}
A=\alpha\frac{1}{f}\frac{x^4}{\bar{x}^3},\qquad B=\beta A\left(\frac{x}{y}\right)^3,
\label{ab}
\end{equation}
and the eccentricity of the orbit
\begin{equation}
e=\sqrt{1-\frac{B^2}{A^2}}=\sqrt{1-\beta^2\frac{x^6}{y^6}}.
\label{eexpr2}
\end{equation}
In the work \cite{Ioketal98} expressions (\ref{ab}) and (\ref{eexpr2}) 
were obtained in the case $\alpha=\beta=1$, which is considered in this article. In \cite{Sasetal16}, the numerical values of the multipliers $\alpha$ and $\beta$ are obtained by solving the equations for the orbit evolution. In the work \cite{Ero18}, the tidal influence of perturbations in dark matter was taken into account and a refined expression for the small semi-axis of the orbit was obtained. The probability distribution for $x$ and $y$ is flat \cite{Naketal97,Sasetal16}
\begin{equation}
dP=\frac{18x^2y^2}{\bar{x}^6}dxdy.
\label{dxdy}
\end{equation}

\begin{figure}
	\begin{center}
\includegraphics[angle=0,width=0.5\textwidth]{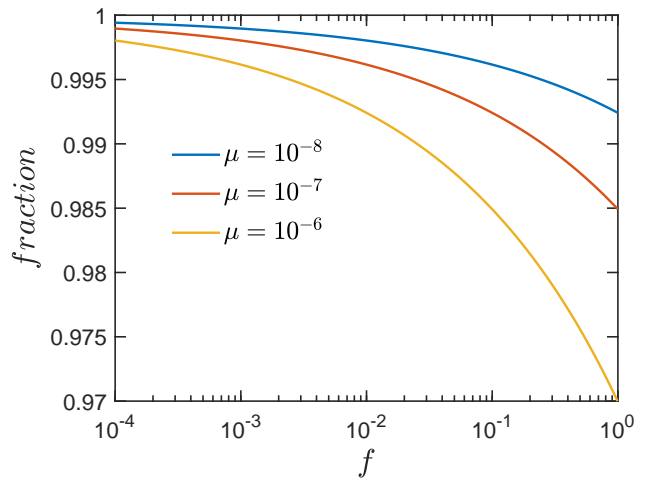}
	\end{center}
\caption{The fraction of clusters that do not intersect during the passage of the periaster, depending on the fraction of PBHs in the dark matter $f$ for different cluster size $\mu=10^{-6}$, $10^{-7}$ and $10^{-8}$ (from down to up).}
	\label{grfra}
\end{figure}

If the gravitational radiation is the only cause of the orbital energy loss, then the lifetime of the pair before the merger \cite{Sasetal16}
\begin{equation}
t_c=\frac{3c^5}{170G^3M_{\rm BH}^3}A^4(1-e^2)^{7/2}.
\label{tc}
\end{equation}
In our case with clusters, there is an additional channel of energy dissipation of orbital motion -- tidal forces, so the distribution over the time of mergers may be different. Let's denote the initial (at the time before the begin of the orbit evolution) ratio of the cluster radius to the average distance 
\begin{equation}
\mu=\frac{R}{\bar x}.
\end{equation}
As shown below, the rate of cluster mergers significantly depends on this value. At the same time, at the moment $t_{\rm eq}$, the ratio $R/A=\mu f/\bar x^4$ takes place. 

The condition that clusters do not intersect each other in the initial period of evolution has the form of a restriction on the periaster of the Kepler orbit $r_p=A(1-e)>2R$. From here we get the constraint on the initial conditions
\begin{equation}
y<x\left[1-\left(1-\frac{2 \mu f}{(x/\bar x)^4}\right)^2\right]^{-1/6}.
\label{uslkas}
\end{equation}
The fraction of clusters that initially satisfy this condition is shown in Fig.~\ref{grfra}. Evolution under the influence of tidal forces follows the path of orbit circularization, so the condition $r_p=A(1-e)>2R$, if it took place at the beginning, usually persists until the clusters touch at the very end of evolution.

\begin{figure}
	\begin{center}
\includegraphics[angle=0,width=0.5\textwidth]{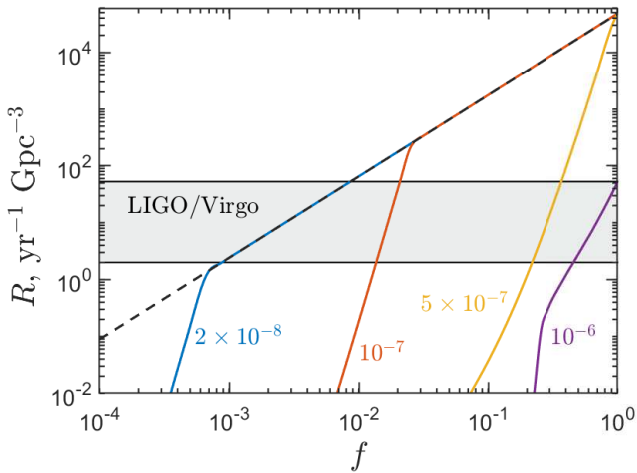}
	\end{center}
\caption{The rate of PBH clusters mergers (\ref{rate1}) depending on their fraction in the dark matter $f$ and the ratio of the cluster radius to the average initial distance between clusters $\mu=R/\bar x$ in the cases (from left to right) $\mu=2 \times 10^{-8}$, $10^{-7}$, $5 \times 10^{-7}$, and $10^{-6}$. At their upper parts the curves converges to the merge rate of the single black holes (dashed line). The shaded area shows the area acceptable from the point of view of LIGO/Virgo observations.}
	\label{gr2cl}
\end{figure}

To calculate the merger rate, it is necessary to determine the area in the space of initial values $A$ and $e$, within which clusters of PBHs merge by the time $t$. To determine this region, the equations describing the evolution of binary (\ref{eqm1}) and (\ref{eqm2}) are solved. Then numerical integration of the area of this region allows one to obtain the probability of $P(<t)$. As a result, we find the rate of gravitational bursts near the present moment of time $t_0$ 
\begin{equation}
R_{gw}=\left.\frac{\rho_c\Omega_mf}{M_{\rm BH}}\frac{dP(<t)}{dt}\right|_{t=t_0},
\label{rate1}
\end{equation}
where $\rho_c=9.3\times10^{-30}$~g~cm$^{-3}$ is the critical density, $\Omega_m\approx0.27$. The result is shown at Fig.~\ref{gr2cl} for the masses of the clusters $M=100M_\odot$. The masses of low-mass PBHs in the clusters have not yet been specified, they will be discussed further. Cluster mergers occur less efficiently than mergers of pairs of single PBHs with the same masses. This is explained by the fact that, on average, the orbit of clusters circulates more efficiently, which increases the time of orbit compression and reduces the rate of cluster mergers.

It can be seen that a decrease in the cluster size naturally leads to a situation of single PBHs, where a head collision is practically excluded, and where tidal forces are small. That is, when the cluster size tends to zero, we come to a situation with single PBHs with the same mass as the cluster. At $M=100M_\odot$, the difference from the case with single PBHs at the level of computational errors is not felt at $\mu\leq10^{-8}$. And at $\mu\geq 10^{-6}$, the rate of mergers becomes less than the observed values even at $f\sim1$. The reason for this is the circularization of the orbit at an early stage of evolution. In this case, the observed merger rate can be explained by the formation of pairs of clusters in the late universe, similar to single PBHs \cite{Bird16, Clesse16} or as a result of direct collision.

An important circumstance is that when $\mu\sim10^{-7}$ the clusters under consideration are very dense. So, for $M=100M_\odot$, their radii are $R\sim2.6\times10^{12}$~cm$=0.17$~AU. The velocity dispersion in clusters is $v\sim717$~cm~s$^{-1}$. The two-body relaxation time in the clusters is
\begin{equation}
\label{tr}
t_{\rm r}\simeq
\frac{v^3}{4\pi G^2m^2n\Lambda_C},
\end{equation}
where $n=M/(mV)$, $V=4\pi R^3/3$, Coulomb logarithm $\Lambda_C\sim15$, and virial velocity
$v\sim (GM/R)^{1/2}$.
In order for the evolution time ($\simeq40t_{\rm r}$) to exceed the Hubble time, it is necessary that small PBHs in the cluster have masses  $m\leq10^{-11}M_\odot$. It is interesting to note that such PBHs fall into the mass window $10^{20}-10^{24}$~g, in which it is possible to explain all dark matter in the form of PBHs \cite{CarKuh20}. Even denser ``extreme'' clusters were considered in \cite{Ero21}. The clusters considered here are structurally close to extreme clusters with collapsed cores. 

Let us now discuss the situation $r_p=A(1-e)<2R$. During the intersection, dynamic friction forces act on the clusters and central BHs. This issue could be studied by direct numerical modelling, which is beyond the scope of this work. For the cluster case, no corresponding simulations have been done yet, but in \cite{KavGagBer18} and \cite{PilTkaIva22}, the interaction of PBHs surrounded by dark matter halos has been studied. Such systems resemble the clusters, so one can expect similar features of evolution. In \cite{KavGagBer18} and \cite{PilTkaIva22}, an important result was obtained that the merger of such PBHs is very effective, and their merger actually occurs at the first intersection. 

Really, the characteristic time of dynamic friction
\begin{equation}
t_{\rm df}\sim \frac{v^3}{4\pi G^2\Lambda B_d\rho M_h},
\end{equation}
where $\Lambda\approx10$, $B_d\approx0.426$. For $M_h=30M_\odot$, $M=100M_\odot$, $\mu\sim10^{-7}$ we get $t_{\rm df}\sim3\times10^{-4}$~years. Thus, the last stage of evolution is going very fast. And even if the clusters intersected only in a very small part of the orbit (in the periastre), they quickly lost orbital energy and merged. 

\begin{figure}
	\begin{center}
\includegraphics[angle=0,width=0.5\textwidth]{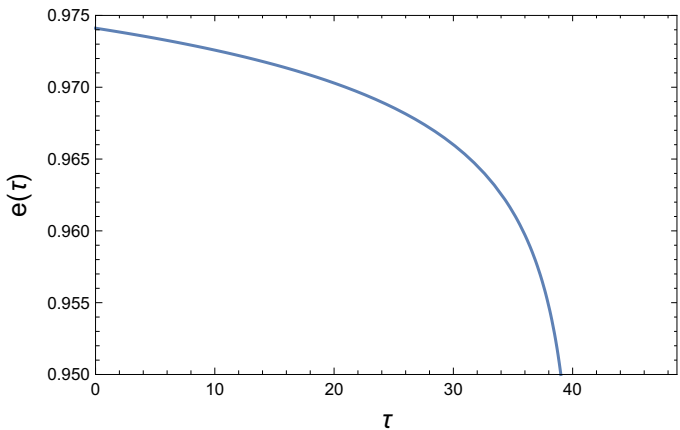}
	\end{center}
\caption{An example of the eccentricity evolution for the binary system's parameters specified in the text. The dimensionless time $\tau$ is equal to the ratio of $t$ and the orbital period in a circular orbit with $a=\bar x$. The current moment of time corresponds to $\tau=935$.}
	\label{ex1}
\end{figure}

Fig.~\ref{ex1} shows one particular example of the evolution of the eccentricity in the case $f=0.01$. The initial values are: $M=100M_\odot$, $\mu=5\times10^{-7}$, $x=0.023$, $y=0.038$, $e=0.974$, $A=10.4$~AU (of the order of Saturn's orbit) and $T=2.4$~years. During the first 590~Myr, the evolution of the binary clusters occurs mainly under the influence of tidal forces. During this time, the semimajor axis decreases by about 20 times, and the eccentricity decreases to $e=0.00008$. At this point, the tidal forces are weak, and losses due to the gravitational radiation begin to prevail. Under the influence of these losses, the orbit is compressed so that the total time of evolution before the merger of clusters is 13.4 billion years, i.e. clusters merge in the modern era and can give contribution to the observed gravitational waves. 

\section{Discussion}

As a rule, it is assumed that mergers of single PBHs in pairs can be responsible for the LIGO/Virgo events. Some studies have considered the mergers of PBHs surrounded by dark matter halos \cite{KavGagBer18,PilTkaIva22}. However, there is a model \cite{RubSakKhl01,KhlRubSak02}, in which clusters are formed rather than single PBHs. In this paper, we investigated the merging of such clusters in order to find out whether in this case it is still possible to explain the LIGO/Virgo signals. As a model system, we took clusters in the center of which there is one massive black hole (with a mass of $\sim30M_\odot$) surrounded by smaller black holes, so that the total mass of the cluster is $\sim100M_\odot$.

Clusters have an additional channel of energy dissipation of orbital motion, which can fundamentally change the result for mergers statistics in comparison with the case of single PBHs. We have found that tidal gravitational interactions lead to rapid circularization of the orbit at the initial stage of evolution. When eccentricity tends to zero, tidal forces also tend to zero and become ineffective. The further evolution of the orbit is determined by the emission of gravitational waves, as in the case of single PBHs. The difference is that this part of the evolutionary track starts from the state of an almost circular orbit, which reduces the efficiency of the radiation of gravitational waves compared to the case of pairs of single PBHs. As a result, it turns out that the rate of cluster pairs merging is suppressed compared to the case of single PBHs, but there remains an allowed range of parameters in which the LIGO/Virgo observations can be explained. 

It is necessary to emphasize the new interesting fact, which is absent in the case of single PBHs, that even for the PBH fraction in the composition of dark matter $f\simeq1$, the merger rate of binary clusters can be sufficiently low and it does not exceed LIGO/Virgo rate. This weakens the constraints on black holes obtained from the LIGO/Virgo observations. If PBHs do not constitute all the dark matter ($f<1$), then the remaining dark matter should form a condensation around the merging clusters, in analogy with single PBH pairs \cite{KavGagBer18, PilTkaIva22}, which at some level affects the rate of mergers. We will consider the corresponding corrections  in the next article.

The LIGO/Virgo statistics can be explained if the ratio of the radius of clusters and the average distances between clusters is $\mu \simeq 5\times10^{-7}$ and the fraction of PBHs in the dark matter is $f\sim 0.1$. The last is consistent with the dynamical and lensing constraints for such massive compact objects \cite{CarKuh20}. Moreover, at $\mu\leq10^{-8}$, the dynamics of a pair of clusters is already indistinguishable from the dynamics of a pair of single PBH. Thus, this model with clusters is significantly limited in the size of clusters. Another strong constrain is the time of internal relaxation of clusters. Due to the fact that the clusters are very dense, relaxation is very effective if the masses of small PBHs are large enough. The 40 relaxation times are comparable to the age of the Universe if the masses of small PBHs are $m\leq10^{-11}M_\odot$. Such masses fall into the mass window, in which the PBHs are able to explain all the dark matter \cite{CarKuh20}. If the PBH masses are greater, then the internal dynamic evolution of the cluster occurs in less than the Hubble time, and this evolution would noticeably influenced by the tidal gravitational disturbances from the second cluster. In future papers we will consider also interesting effects associated with the internal two-body evolution of clusters and corresponding mass loss. 

In this article, we limited ourselves only to those clusters that could contribute to the LIGO/Virgo gravitational wave signals. However, the theoretical model \cite{RubSakKhl01,KhlRubSak02} allows for the existence of clusters in very wide parameter ranges, both in terms of the masses of constituent PBHs, and the sizes and total masses of the clusters themselves. Therefore, the existence of clusters with significant internal two-body dynamic evolution is not excluded. When applied to binary clusters, tidal interactions will lead to the heating of the clusters (an increase in the internal PBHs velocity dispersion), which will be accompanied by the flow (dynamic evaporation) of the PBHs from the cluster. The reason for this is the growth of the Maxwell tail of the velocity distribution after each tidal impact. This restructuring of the cluster leads to the flow of objects from it, which will be discussed in detail in the future articles. The results obtained can also be applied to the process of merging of two globular star clusters, since the effect of tidal heating is similar in basic respects. As far as we know, binary globular star clusters have not yet been observed, but their existence is not excluded in stellar bulges or compact dense spheroid galaxies. 

\bigskip

This work is supported by the Russian Science Foundation grant 23-22-00013, 

https://rscf.ru/en/project/23-22-00013/.


\begin{thebibliography}{999}


\bibitem{ZelNov66}  Zel’dovich, Y.B.;  Novikov I.D. The Hypothesis of Cores Retarded during Expansion and the Hot Cosmological Model. {\em Soviet Astronomy} {\bf 1967}, {\em 10}, 602-603. 

\bibitem{Haw71} Hawking, S. Gravitationally collapsed objects of very low mass, {\em Monthly Notices of the Royal Astronomical Society} {\bf 1971}, {\em 152}, 75.

\bibitem{Car75} Carr, B. J. The primordial black hole mass spectrum, {\em  The Astrophysical Journal} {\bf 1975}, {\em  201}, 1-19.

\bibitem{KhlPol0} Khlopov, M. Yu; Polnarev, A. G. Primordial black holes as a cosmological test of grand unification. {\em Physics Letters} {\bf 1980}, {\em  B 97},  383.

\bibitem{BerKuzTka83} Berezin, V. A.; Kuzmin, V. A.; Tkachev, I. I. Thin-wall vacuum domain evolution.  { \em Physics Letters} {\bf 1983},  {\em  B 120}, 91.

\bibitem{RubSakKhl01}  Rubin, S.G.;  Sakharov, A.S.;  Khlopov, M.Yu. The Formation of Primary Galactic Nuclei during Phase Transitions in the Early Universe. {\em Journal of Experimental and Theoretical Physics} {\bf 2001}, {\em 91},  921-929; arXiv:hep-ph/0106187.

\bibitem{KhlRubSak02} Khlopov, M.Yu.; Rubin, S.G.;  Sakharov A.S. Strong primordial
inhomogeneities and galaxy formation. {\em Gravitation and Cosmology} {\bf 2002}, {\em 8}, 57–65; arXiv:astro-ph/0202505.

\bibitem{DolSil93} Dolgov, A.; Silk, J. Baryon isocurvature fluctuations at small scales and baryonic dark matter.
\emph{Phys. Rev.} {\bf 1993}, {\em D 47},  4244.

\bibitem{Caretal20} Carr, B.; Kohri, K.; Sendouda, Y.; Yokoyama J.
Constraints on Primordial Black Holes. arXiv:2002.12778.

\bibitem{CarKuh20} Carr, B. ; Kühnel, F. Primordial Black Holes as Dark Matter: Recent Developments. {\em Annual Review of Nuclear and Particle Science} {\bf 2020}, {\em 70}, 355-394 (2020); arXiv:2006.02838.

\bibitem{Naketal97}  Nakamura, T.;  Sasaki, M.;  Tanaka, T.;  Thorne, K.S. Gravitational Waves from Coalescing Black Hole MACHO Binaries.
{\em  The Astrophysical Journal} {\bf 1997}, {\em  487}, L139.

\bibitem{Ioketal98}  Ioka, K.;  Chiba, T.;  Tanaka, T.;  Nakamura, T.
Black Hole Binary Formation in the Expanding Universe -- Three Body Problem Approximation. {\em  Physical Review} D {\bf 1998}, {\em 58},  063003; arXiv:astro-ph/9807018.

\bibitem{Sasetal16}  Sasaki, M.;  Suyama, T.;  Tanaka, T.; Yokoyama S. Primordial black hole scenario for the gravitational wave event GW150914. {\em  Physical Review Letters} {\bf 2016}, {\em  117}, 061101; arXiv:1603.08338.

\bibitem{GneOst99} Gnedin, O.Y.; Ostriker, J.P. On the Self-consistent Response of Stellar Systems to Gravitational Shocks. {\em The Astrophysical Journal} {\bf 1999}, {\em 513}, 626-637 (1999); arXiv:astro-ph/9902326.

\bibitem{Gneetal99}  Gnedin, O.Y.; Hernquist, L.; Ostriker, J.P. Tidal Shocking by Extended Mass Distributions. {\em The Astrophysical Journal} {\bf 1999}, {\em 514}, 109-118 (1999).

\bibitem{Raidal18} Raidal, M.; Spethmann, C.; Vaskonen, V.; Veermae, H. Formation and Evolution of Primordial Black Hole Binaries in the Early Universe. {\em Journal of Cosmology and Astroparticle Physics} {\bf 2019}, {\em 02}, 018; 
1812.01930 [astro-ph.CO]

\bibitem{Vaskonen19} Vaskonen, V.; Veermae, H. Lower bound on the primordial black hole merger rate. {\em Physical Review D} {\bf 2020}, {\em 101}, 043015; 
arXiv:1908.09752 [astro-ph.CO]

\bibitem{Jedamzik20} Jedamzik, K. Primordial Black Hole Dark Matter and the LIGO/Virgo observations. {\em Journal of Cosmology and Astroparticle Physics} {\bf 2020}, {\em 09}, 022; 
2006.11172 [astro-ph.CO]

\bibitem{KavGagBer18} Kavanagh, B.J., Gaggero D., Bertone, G. Black Holes' Dark Dress: On the merger rate of a subdominant population of primordial black holes. {\em Physical Review D} {\bf 2018}, {\em 98}, 023536; arXiv:1805.09034 [astro-ph.CO]

\bibitem{PilTkaIva22} Pilipenko, S.; Tkachev, M.; Ivanov, P.
The evolution of a primordial binary black hole due to interaction with cold dark matter and the formation rate of gravitational wave events. arXiv:2205.10792.

\bibitem{Ero21} Eroshenko, Yu. Mergers of primordial black holes in extreme clusters and the $H_0$ tension. {\em Physics of the Dark Universe} {\bf 2021}, {\em 32}, 100833; arXiv:2105.03704.

\bibitem{LL-2} Landau, L.D.; Lifshitz, E.M. {\em The Classical Theory of Fields}; Butterworth-Heinemann: Oxford, United Kingdom, 1980.

\bibitem{LL-1} Landau, L.D.; Lifshitz, {\em Mechanics}; Butterworth-Heinemann: Oxford,	United Kingdom, 1976.

\bibitem{Ero18}  Eroshenko, Yu.N. Gravitational waves from primordial black holes collisions in binary systems. {\em Journal of Physics: Conference Series} {\bf 2018}, {\em 1051}, 012010; arXiv:1604.04932.

\bibitem{Bird16}
Bird, S.; Cholis I.; Mu\~n{}oz, J.B.; Ali-Ha\"\i{}moud, Y.; Kamionkowski, M.; Kovetz, E.D.; Raccanelli, A.; Riess, A.G. Did LIGO detect dark matter?
{\em Physical Review Letters} {\bf 2016}, {\em 116}, 201301
arXiv:1603.00464 [astro-ph.CO].

\bibitem{Clesse16}
Clesse, S.; Garc\'\i{}a-Bellido, J. The clustering of massive Primordial Black Holes as Dark Matter: measuring their mass distribution with Advanced LIGO
{\em Physics of the Dark Universe} {\bf 2017}, {\em 15}, 142-147
1603.05234 [astro-ph.CO].


\end{thebibliography}
\end{document}